# Teaching assistants' performance at identifying common introductory student difficulties revealed by the conceptual survey of electricity and magnetism

Nafis I. Karim[1], Alexandru Maries[2], and Chandralekha Singh[1]

[1] *Department of Physics and Astronomy, University of Pittsburgh, Pittsburgh, PA 15260*
[2] *Department of Physics, University of Cincinnati, Cincinnati, OH 45221*

We discuss research involving the Conceptual Survey of Electricity and Magnetism (CSEM) to evaluate one aspect of the pedagogical content knowledge of teaching assistants (TAs): the knowledge of introductory students' alternate conceptions in electricity and magnetism as revealed by the CSEM. For each item on the CSEM, the TAs were asked to (1) identify the most common incorrect answer choice of introductory physics students and (2) predict the percentage of introductory students who would answer the question correctly in a post-test. Then, we used CSEM post-test data from approximately 400 introductory physics students (provided in the original paper describing the CSEM) to assess the extent to which the TAs were able to identify the alternate conceptions of introductory students related to electricity and magnetism. In addition, we conducted think-aloud interviews with TAs who had at least two semesters of teaching experience in recitations to explore their reasoning about this task. We find that the TAs struggled to think about the difficulty of the questions from introductory students' perspective and they often underestimated the difficulty of the questions. Moreover, the TAs often expected certain incorrect answer choices to be common among introductory students when in fact those answer choices were not common.

## I. INTRODUCTION

The Conceptual Survey of Electricity and Magnetism (CSEM) is a conceptual multiple-choice survey [1] commonly used to assess student learning in introductory electricity and magnetism courses. Pedagogical content knowledge, or PCK as defined by Shulman [2], includes "Understanding of the conceptions and preconceptions that students bring with them to the learning of those most frequently taught topics and lessons." According to this definition, knowledge of students' common alternate conceptions is one aspect of PCK [2]. The research presented here uses the CSEM to explore this aspect of the PCK of physics graduate Teaching Assistants (TAs) in the context of electricity and magnetism. In particular, we investigate the extent to which physics TAs are able to identify common alternate conceptions of introductory students on individual items on the CSEM. Knowledge of the common difficulties and of the types of reasoning used by introductory physics students can be helpful in designing pedagogical strategies to improve student learning [2-6].

## II. METHODOLOGY

The study was carried out over three years with 81 first year physics graduate TAs enrolled in a mandatory semester long pedagogy oriented TA training class at the University of Pittsburgh (Pitt), which meets once a week for two hours. The recitations and labs were mostly taught by these TAs in a traditional manner. The TA training course is a general introduction to pedagogical issues in physics teaching and learning. It focuses on helping TAs become more effective teachers and was not tailored specifically to the TA's teaching assignments that semester. The TAs participated in this study during their first semester and thus, they had limited teaching experience in introductory physics recitations and labs.

In all years, for each CSEM item, the TAs identified what they expected to be the most common incorrect answer choice of introductory students. We refer to this task as the CSEM-related PCK task. In years two and three of the study, the researchers also asked TAs to predict the percentage of students who would answer each CSEM question correctly in a post-test (after traditional instruction in relevant concepts). In addition, 11 TAs participated in think-aloud interviews [7] during which they identified what they expect to be the most common incorrect answer choice of introductory students while thinking out loud. When interviewed, each of these 11 TAs had more than two semesters of teaching experience in recitations. The introductory student data used to analyze TA performance is from Ref. [1].

## III. RESULTS

Many TAs noted that the task of thinking from an introductory physics student's point of view was challenging; some even confessed that they did not feel confident about their performance in identifying the most common incorrect answers. Although there are 32 questions on the CSEM, due to space constraints, we focus only on a few insightful items relevant to the research questions RQ1 and RQ2 discussed below:

**RQ1** Are there situations in which a significant fraction of TAs selects answer choices that very few introductory students select? What are some common examples of reasoning that TAs use to select those answer choices?

**RQ2** To what extent are TAs able to predict the difficulty of the questions?

Table I shows the percentages of introductory algebra-based students who selected each answer choice from most to least common after traditional instruction as reported in Ref. [1] as well as the percentages of TAs who selected these choices (the TAs were asked to identify the most common incorrect answer choices of introductory students). The usage of red font is to single out the common alternate conceptions held by 20% or more of introductory students as given in Ref. [1]. We note that in all that follows, "students" refers to introductory algebra-based physics students.

TABLE I. Selected questions (Q # refers to item number) on the CSEM, percentages of introductory algebra-based physics students who answered the questions correctly in a post-test, percentages of introductory students who selected each incorrect answer choice ranked from most to least common (asked to select the correct answers), the most common incorrect answer choices selected by TAs (asked to select the most common incorrect answer if students did not know the correct answer after traditional instruction). The red font is used to identify common student alternate conceptions (i.e., incorrect answer choices selected by 20% or more introductory students).

| Q # | Correct Answer | Students' incorrect choices | | | | TAs' incorrect choices | | | |
|---|---|---|---|---|---|---|---|---|---|
| | | 1st | 2nd | 3rd | 4th | 1st | 2nd | 3rd | 4th |
| 1 | 63 B | 23 C | 7 D | 4 A | 3 E | 54 C | 19 D | 15 E | 11 A |
| 2 | 42 A | 21 B | 19 E | 11 D | 5 C | 49 D | 25 B | 16 E | 10 C |
| 13 | 51 E | 27 A | 20 B | 1 C | 0 D | 56 A | 42 B | 1 C | 1 D |
| 14 | 16 D | 54 A | 13 E | 9 B | 4 C | 46 A | 24 E | 18 B | 11 C |
| 26 | 49 A | 21 D | 11 B | 6 C | 6 E | 47 C | 29 D | 17 B | 7 E |
| 28 | 40 C | 35 E | 12 B | 8 A | 3 D | 55 E | 32 A | 11 B | 2 D |

*Charge distribution on conductors/insulators (Q1, Q2)*

Q1 and Q2 on the CSEM ask about what happens to an excess charge placed at some point P on a conducting (Q1) or insulating hollow sphere (Q2). For Q1, students' most common alternate conception (23% of students) was that the charge distributes evenly on the inside and outside of the metal sphere (answer choice C). For Q2, there were two alternate conceptions: that the charge distributes itself evenly on the outside of the sphere (i.e., not distinguishing between insulating and conducting – answer choice B selected by 21% of students) and that there will be no excess charge left (answer choice E, selected by 19% of students). For Q2, many TAs expected that the most common incorrect answer selected by students would be answer choice D (which states that most of the charge remains at point P, but some of it will spread over the sphere). Answer choice D was selected by 49% of the TAs as the most common incorrect answer choice of students, but only 11% of students actually selected this choice according to Ref. [1].

For Q1, some of the TAs reasoned that D would be the most common incorrect answer choice because students would expect that the charges would move, but that there is not enough force to move all the charges everywhere around the sphere, or that it takes more than a few seconds for the charge to spread everywhere and therefore some will remain at point P. The two quotes below exemplify the reasoning two TAs used: "They [students] don't expect that for a metal [there is] enough push in order to move all the charges from [P]." "Most people probably think it's D […] because they might not recognize that it has to be an instantaneous distribution of charge. So they recognize that the charge will have to spread over the surface, and since we know it's metal, I'm assuming they understand a conductor won't have charge on the inside. It [charge] is all gonna be on the surface, but they might assume that the majority of the charge hasn't fully distributed yet."

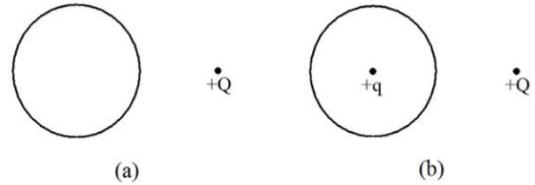

FIG 1. Diagrams for Q13 (a) and Q14 (b) on the CSEM

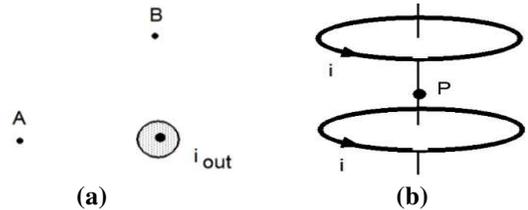

FIG. 2. Diagrams for Q26 (a) and Q28 (b) on the CSEM

For Q2, the most common reasoning of TAs for selecting answer choice D as the most common incorrect answer choice of students was that it was the incorrect answer choice that is most similar to the correct answer and that students may have some understanding that an insulating sphere is different from a conducting sphere, but may not fully understand it. For example, one TA stated: "They will choose D because they know something about insulating that it is not like the conducting, but they [may not know] that the charge will stay at the position [P]."

*Induced charge and electric field/force (Q13, Q14)*

The diagrams provided for CSEM items Q13 and Q14 are shown in Fig. 1. In Q13, the sphere is hollow and conducting and has an excess positive charge on its surface. The question asks for the direction of the electric field at the center of the sphere. In Q14, the sphere is also hollow and conducting, but it has no excess charge, and the question asks about the forces acting on the two charges. On both of these questions, the most common difficulty of students is to not recognize that the conducting sphere alters the electric field/forces. Thus, in Q13, 27% of them selected choice A in which the electric field is to the left (as though the sphere does not affect it) and in Q14, 54% of them also selected choice A in which the forces that the two charges feel are the same (once again, as though the sphere does not affect the forces) [1]. While the TAs did very well at identifying the common alternate conceptions in Q13, on Q14, 53% of the TAs selected answer choices B, C, and E which combined were selected by only 26% of students [1].

In the interviews, the majority of the TAs selected answer choice A as the most common incorrect answer choice on Q13 because they recognized that students may have the difficulty mentioned above (ignore the sphere), but on Q14, the TAs often selected a different difficulty, typically that students would expect the charge spread on the sphere to "do something", i.e. the TAs' responses were guided by different reasoning when selecting the most common incorrect answer choice for Q14. For example, on Q13, a TA stated: "Maybe someone would say leftward because they think of the positive being the source so they think of it making a [field] line and the [field] line is going outward from the charge, and they think it's just going to go straight through the sphere." But on Q14 the same TA thought that students are most likely to select answer choice E (both charges feel a force, but the forces are different) and reasoned that: "They might think that little q at the center of the sphere […] is feeling forces from the charges that are distributed along the surface [of the sphere], and big Q here might feel force from this guy [q] and all the surface charges [on the sphere]."

Another TA also selected answer choice A in Q13 by citing similar reasoning (students will not know that the sphere has an effect), and when answering Q14 she initially selected answer choice A. However, after noticing answer choice E, she changed her mind and stated: "I think E [may be most common] because they might realize that the sphere does do something to change things, so they think 'ok, I know [the forces would normally be] equal and opposite, but now there's a sphere here, so I don't know exactly how that works' [i.e., what the effect of the sphere is] so they'll just throw in something [i.e., include some effect due to the sphere], so E is that something." It appears that this TA was aware that students may be guided by similar incorrect thinking (conducting sphere will not have an effect) in Q14 as in Q13, but in Q14, selected the answer choice which incorporates a correct idea (conducting sphere has an effect), but is missing another idea in order to be correct.

In many other questions, TAs often selected answer choices not common among students for similar reasons. For example, in Q2, some TAs thought that students would think that most of the charge stays where it is put, but that some of the charge does spread over the sphere, which is an answer choice that incorporates a partially correct idea. The TAs sometimes explicitly noted that they selected this answer choice as the most common incorrect answer choice because it is the one that is most similar to the correct answer. While sometimes using this strategy to identify the most common incorrect answer choice led to prediction of students' difficulty correctly, it often misled the TAs into selecting an answer choice that was not very common. For example, on Q14, this reasoning led some TAs to select choice E as the most common incorrect answer of students, an answer choice selected only by 13% of students.

*Magnetic field caused by a current (Q23, Q26, Q28)*

On Q23, there were no common alternate conceptions (i.e., no answer choices selected by 20% or more students). Q26 provides the diagram shown in Fig. 2 (a) and asks students to identify the direction of the magnetic field at points A and B. On this question, the most common alternate conception of students is that the magnetic field is radially outward from the wire (answer choice D, selected by 21% of students). However, 47% of the TAs selected answer choice C in which the direction of the magnetic field is opposite to the correct direction (i.e., clockwise instead of counterclockwise) and only 6% of students selected this answer choice. This resulted in a low average PCK score on this question (35%). All the interviewed TAs who selected this answer choice essentially stated that students may either use their left hand or use the right hand rule incorrectly, but the choices selected by many students do not suggest this as a major difficulty.

For Q26, in Fig. 2 (a), some interviewed TAs used similar reasoning as some of the TAs who selected answer choice E in Q14 – students have some correct ideas (try to use the right hand rule but obtain the incorrect direction). It is important to point out that after recognizing that students may be answering the question incorrectly for this reason (which does not seem to be common), the interviewed TAs often did not consider all the other answer choices carefully, and did not realize that students may have other alternate conceptions, namely that the magnetic field would be radially outward from the wire (i.e., confusion between electric and magnetic field). After the TAs answered all the other questions in the interview, they were often asked to return to this question and think about whether they expected that any students would select answer choice D (radially outward magnetic field). After being asked to consider this answer choice explicitly, they were often able to recognize the alternate conception guiding students to select answer choice D and some interviewed TAs wanted to change their original answer. Similar to Q14, some TAs attempted to identify common alternate conceptions for Q26 by arguing that students may have some correct ideas, but miss something that causes them to not have the fully correct answer. However, it appears that for this question (and others mentioned earlier), this type of reasoning from the TAs often steered them in the wrong direction and led them to identify an answer choice that is not common among students while missing the most common alternate conception.

Q28 provides students with the diagram shown in Fig. 2 (b) and asks for the direction of the magnetic field at point P. Here, the most common alternate conception of students is that the two magnetic fields created by the two wires cancel out (answer choice E selected by 35% of students). Here, the majority of TAs (55%) selected this answer choice, but 32% of them selected choice A (downward) which only 8% of students selected. Similarly to Q26 discussed above, all of the TAs who selected this answer choice during interviews claimed that students may use the right hand rule incorrectly and obtain the incorrect direction. However, it appears that very few students do this.

Figure 3 shows TAs' average predictions of the difficulty of each question on the CSEM, i.e., TAs' predictions of the percentage of students who answered each question correctly as well as the actual difficulty of each question (National Data from Ref. [1]). Figure 3 shows that the TAs underestimated the difficulty of the majority of the questions on the CSEM. The discrepancy between TAs' predicted difficulty and the actual difficulty is quite large for some questions, in particular, for the questions that were most difficult for students (e.g., items 14, 20, 24, 29, 31, 32). Figure 3 also shows that TAs' predicted difficulty does not fluctuate very much. With the exception of only five questions, the TAs' predicted difficulty is between 45% and 65%, thus indicating that the TAs did not have a good sense of how difficult the questions are from the perspective of students. This conclusion is further supported by averaging TAs' predictions over all questions and comparing them to the actual difficulty: TAs over-predicted students' performance on the CSEM by 15% on average.

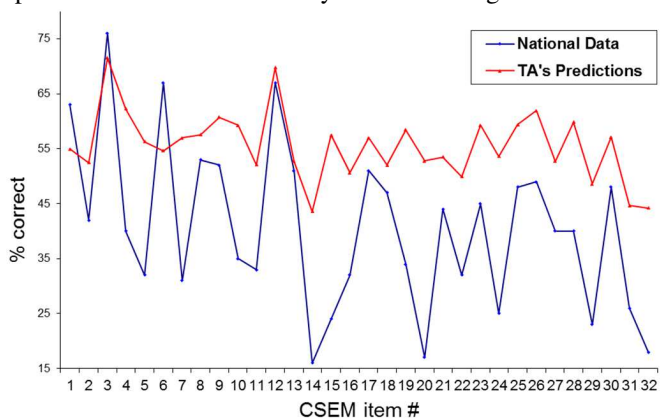

FIG 3. Comparison of the percentages of correct student responses to each question predicted by the TAs (i.e., TA predictions of the item difficulty of all questions) shown in red with that of students' actual performance on each question from Ref. [1] shown in blue. Standard Deviations of TA data range from 17.7 to 24.6 (not shown for clarity).

## IV. SUMMARY

Our investigation used the CSEM to investigate the extent to which physics TAs are aware of common student alternate conceptions of electricity and magnetism. For each item on the CSEM, the TAs were asked to identify what they thought was the most common incorrect answer choice of students after traditional instruction. Think-aloud interviews were conducted with 11 TAs who had more than two semesters of teaching experience in recitations to obtain an in-depth account of what reasoning TAs used to predict that certain alternate conceptions may be most common. Additionally, in years two and three of the study, the TAs were also asked to estimate the difficulty of each question.

We found that when trying to decide what answer choices would be common among students, TAs often selected answer choices which incorporate both correct and incorrect ideas. While this approach was sometimes productive in helping them identify the most common incorrect answer choice of students, it often led to TAs selecting answer choices that were not very common. Equally importantly, after TAs identified a particular answer choice which incorporated a correct and incorrect idea, they often neglected to consider other answer choices carefully or think about what alternate conceptions could lead to students selecting them. This suggests that a productive approach to helping TAs identify common incorrect answer choices of students may be to explicitly ask them to first identify what alternate conceptions may lead students to select each incorrect answer choice first (for a particular question) and only afterwards ask them to decide which one they expected to be most common.

We also found that the TAs typically underestimated the difficulty of the questions on the CSEM, especially on the challenging questions. For all but five questions, TAs' average predictions for the percentage of students who answered the questions correctly are between 45% and 65%, while the actual percentages vary much more widely. This finding suggests that the TAs struggled to think about the difficulty of the questions from a student's perspective.

In the think-aloud interviews, many TAs explicitly noted that the CSEM-related PCK task was challenging and it was difficult for them to think about physics questions from a student's perspective (e.g., "I don't know introductory students well enough…"). However, many TAs noted that the CSEM-related PCK task was worthwhile and helped them think about the importance of putting themselves in their students' shoes in order for teaching and learning to be effective, especially after receiving data on how students actually performed and discussing particular alternate conceptions. The findings of this investigation can be useful to improve the professional development of physics TAs.

[1] D. P. Maloney, T. L. O'Kuma, C. J. Hieggelke and A. Van Heuvelen, Am. J. Phys. **69**, s12 (2001).
[2] L. S. Shulman, Harvard Educ. Rev. **57**, 1 (1987).
[3] P. M. Sadler, G. Sonnert, H. P. Coyle, N. Cook-Smith and J. L. Miller, Am. Educ. Res. J. **50** (5) 1020 (2013).
[4] A. Maries and C. Singh, Phys. Rev. Special Topics Phys. Educ. Res. **9**, 020120 (2013).
[5] A. Maries and C. Singh, 2014 Physics Education Research Conference Proceedings, p. 171 (2015) http://dx.doi.org/10.1119/perc.2014.pr.039
[6] A. Maries and C. Singh, Phys. Rev. Special Topics Phys. Educ. Res. **12**, 010131 (2016).
[7] K. Ericsson & H. Simon, Psych. Rev. **87**, 215 (1980).